\documentclass{pkas}

\def\year{2014} % publication year
\def\month{March} % publication month
\def\volume{00} % journal volume
\def\issue{0} % journal issue
\def\beginpage{1} % first page of article
\def\endpage{99} % last page of article
\def\received{---} % date paper was received by PKAS
\def\accepted{---} % date of acceptance
\date{Received \received ; accepted \accepted}

\newcommand\ion[2]{{#1}\,{\sc #2}} % ions: \ion{C}{iv} = C IV

\title{
Overview of North Ecliptic Pole Deep multi-wavelength Survey (NEP-Deep)
}

\author[1,18]{H.~Matsuhara}
\author[1]{T.~Wada}
\author[1]{N.~Oi}
\author[1]{T.~Takagi}
\author[1]{T.~Nakagawa}
\author[1,18]{K.~Murata}
\author[2]{T.~Goto}
\author[3]{S.~Oyabu}
\author[3]{T.T.~Takeuchi}
\author[3]{K. Ma{\l}ek}
\author[3]{A. Solarz}
\author[4]{Y.~Ohyama}
\author[5,7]{T.~Miyaji}
\author[6,7,17]{M.~Krumpe}
\author[8]{H.M.~Lee}
\author[8]{M.~Im}
\author[9]{S.~Serjeant}
\author[9,10,16]{C.P.~Pearson}
\author[9,10]{G.J.~White}
\author[11]{M.A.~Malkan}
\author[12]{H.~Hanami}
\author[12]{T.~Ishigaki}
\author[13]{D.~Burgarella}
\author[13]{V.~Buat}
\author[14,15]{A. Pollo}

\affil[1]{ Institute of Space and Astronautical Science, Japan Aerospace Exploration Agency, Sagamihara, Kanagawa 229-8510, Japan ; \email{maruma@ir.isas.jaxa.jp, wada@ir.isas.jaxa.jp, nagisaoi@ir.isas.jaxa.jp, takagi@ir.isas.jaxa.jp, nakagawa@ir.isas.jaxa.jp}}
\affil[2]{ Institute of Astronomy, National Tsing Hua University,  Hsinchu, Taiwan 30013, R.O.C; \email{tomo@phys.nthu.edu.tw}}
\affil[3]{ Graduate School of Science, Nagoya University, Nagoya, Aichi 464-8602, Japan; \email{oyabu@u.phys.nagoya-u.ac.jp, takeuchi@iar.nagoya-u.ac.jp, malek.kasia@gmail.com, quotidi4n@gmail.com}}
\affil[4]{ Academia Sinica, Institute of Astronomy and Astrophysics, Taiwan; \email{ohyama@asiaa.sinica.edu.tw}}
\affil[5]{ Instituto de Astronomia, Universidad Nacional Autonoma de Mexico, Ensenada, Baja California, Mexico; \email{miyaji@astrosen.unam.mx}}
\affil[6]{ Max-Planck-Institut f{\"u}r extraterrestrische Physik, Giessenbachstra\ss{\`e}, 85748 Garching, Germany}
\affil[7]{ University of California, San Diego, Center for Astrophysics and Space Sciences, La Jolla, CA, USA; \email{mkrumpe@ucsd.edu}}
\affil[8]{ Department of Physics \& Astronomy, FPRD, Seoul National University, Seoul 151-742, Korea; \email{hmlee@astro.snu.ac.kr, mim@astro.snu.ac.kr}}
\affil[9]{ Department of Physical Sciences, The Open University, Milton Keynes, MK7 6AA, UK; \email{S.Serjeant@open.ac.uk, glenn.white@open.ac.uk}}
\affil[10]{ Rutherford Appleton Laboratory Oxon, OX11 0QX, UK; \email{chris.pearson@stfc.ac.uk}}
\affil[11]{ University of California, Los Angeles, CA 90095-1547, USA; \email{malkan@astro.ucla.edu}}
\affil[12]{ Iwate University, 3-18-34 Ueda, Morioka, Iwate 020-8550, Japan; \email{hanami@iwate-u.ac.jp, ishigaki@iwate-u.ac.jp}}
\affil[13]{ Aix-Marseille Universit{\'e}, CNRS, LAM, UMR7326, 13388, Marseille, France; \email{denis.burgarella@oamp.fr, veronique.buat@oamp.fr}}
\affil[14]{National Center for Nuclear Research, Poland; \email{Agnieszka.Pollo@fuw.edu.pl}}
\affil[15]{Jagiellonian University Observatory, Poland }
\affil[16]{Oxford Astrophysics, Denys Wilkinson Building, University of Oxford, Keble Rd, Oxford OX1 3RH, UK }
\affil[17]{ESO Headquarters, Karl-Schwarzschild-Str. 2, 85748 Garching, Germany }
\affil[18]{ Department of Space and Astronautical Science, The Graduate University for Advanced Studies, Japan }
\def\corrauthor{
H. Matsuhara
}

\def\runningauthor{
Matsuhara et al.
}

\def\runningtitle{
NEP Deep Survey
}

\def\keywords{
infrared: galaxies --- galaxies: starburst --- AGN --- dust
}

\def\abstracttext{
The recent updates of the North Ecliptic Pole deep (0.5~deg$^2$, NEP-Deep)
multi-wavelength survey covering from X-ray to radio-wave is presented.
The NEP-Deep provides us with several thousands of 15~$\mu$m or 18~$\mu$m 
selected sample of galaxies, which is the largest sample ever made at this 
wavelengths. A continuous filter coverage in the mid-infrared wavelength
 (7, 9, 11, 15, 18, and 24~$\mu$m) is unique and vital to diagnose the 
contributions from starbursts and AGNs in the galaxies out to $z$=2.
The new goal of the project is to resolve the nature of the cosmic star formation
history at the violent epoch (e.g. $z$=1--2), and to find a clue to understand
its decline from $z$=1 to present 
universe by utilizing the unique power of the multiwavelength survey. The progress
in this context is briefly mentioned.
}

\begin{document}
\pkashead %% set title, authors, abstract, etc.

%%%%%%%%%%%%%%%%%%%%%%%%%%%%%%%%%%%%%%%%%%%%%%%%%%%%%%%%%%%%%%%%%%%%%
%%% BEGIN MAIN TEXT HERE %%%%%%%%%%%%%%%%%%%%%%%%%%%%%%%%%%%%%%%%%%%%
%%%%%%%%%%%%%%%%%%%%%%%%%%%%%%%%%%%%%%%%%%%%%%%%%%%%%%%%%%%%%%%%%%%%%

\section{Introduction}

The mid- and far-infrared (MIR and FIR) wavelengths are quite important probes to 
explore the star formation and growth of super-massive black holes (SMBHs) in the 
universe, since the early stages of the star-formation and interacting processes with
active galactic nuclei (AGN) most likely take place within nuclear regions obscured 
by dust. The North Ecliptic Pole (NEP) survey~\citep{matsuhara2006} was one of the 
large area surveys of the AKARI/IRC, and is optimally designed to explore the dust
obscured universe up to $z \sim 2$, with unique, unpararelled continous wavelength coverage
over the 8-24~$\mu$m wavelength gap of the {\it Spitzer}, namely by the existence 
of 9, 11, 15, and 18~$\mu$m bands. The AKARI NEP survey consists of two survey 
projects; deeper one is `NEP-Deep'~\citep[0.5~deg$^2$,][]{wada2008} while a shallower but wider
one is `NEP-Wide'~\citep[5.4~deg$^2$,][]{lee2009}. In order to effectively utilize the value 
of unique MIR data of AKARI, numerous multiwavelegth data, from X-ray to radio-wave, 
were obtained. The status of multi-wavelength data available for NEP-Deep at the time of
the 2nd AKARI conference was presented in \citet{matsuhara2012}. This paper aims to 
describe the major updates over the last few years.

\begin{figure*}[t]
\centering
\includegraphics[width=150mm]{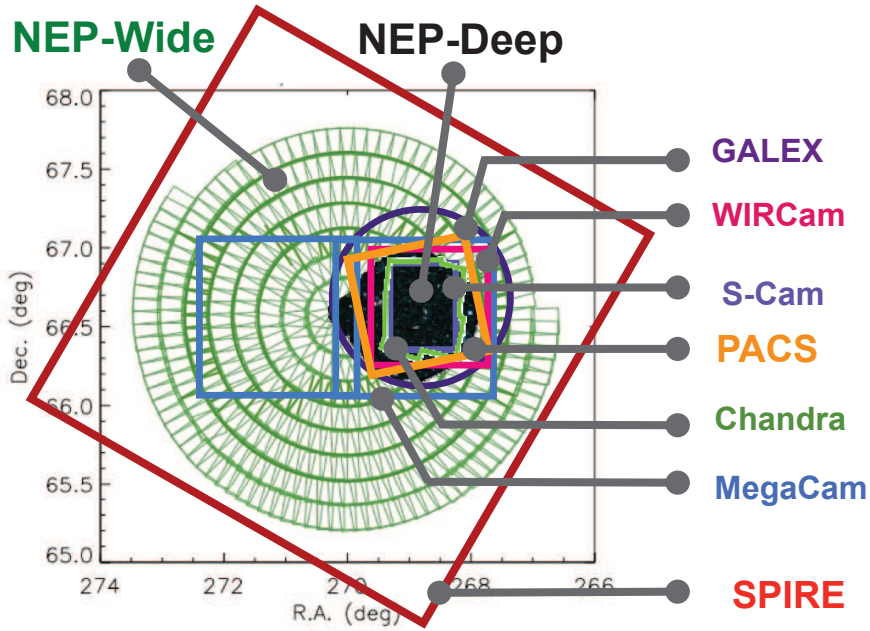}
\caption{Sky coverage of the multi-wavelength data around NEP.  See Table~\ref{tab:maruma1} 
for more information (e.g. areas, depths).   New optical-NIR data is also shown in 
Figure~\ref{fig:maruma2}. \label{fig:maruma1}}
\end{figure*}

It is noteworthy to put a few sentences on the science goal of this mutltiwavelength 
survey projects. One of our major science goals was ``to reveal the cosmic star-formation
 (CSF) history at $z$=1--2", and this has been already achieved to some extent: in case of 
AKARI, \citet{goto2010} presented. After the launch of Herschel the understanding
of CSF history, based on other multiwavelength survey projects (GOODS, COSMOS, etc.) has been
extended to $z \sim 4$~\citep{burg2013,madi2014}. Therefore, we 
now set the new goal as ``to resolve the nature of CSF at the violent epoch (e.g. 
$z$=1--2), and to find a clue to understand the decline of CSF from $z$=1 to present 
universe", by using the power of the  NEP multiwavelength survey data: for example, we can 
classify the dusty AGN and starburst from the MIR spectral energy distributions (SEDs), 
evaluate the star-formation strength (starburstiness) with the ratio of total IR luminosity 
and rest 8~$\mu$m luminosity~\citep{elbaz2011}, and estimate the dust attenuation from 
UV to FIR SED fitting. In section~\ref{sec:outcome}, we also briefly highlight them.

\section{Recent Progress in Multiwavelength Data\label{sec:updates}}

\begin{figure}[h]
\centering
\includegraphics[width=80mm]{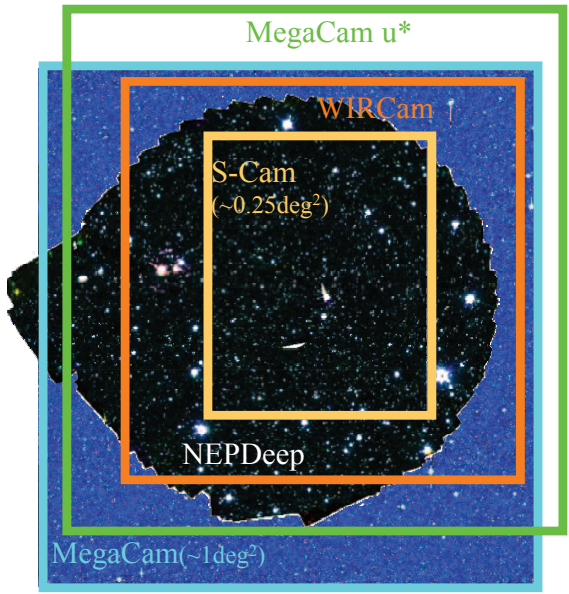}
\caption{Sky coverage of optical-NIR images recently obtained with CFHT~\citep{oi2014}: WIRCAM 
$Y$, $J$, $K_{\rm s}$ and MegaCam $u^{*}$, $g'$, $r'$, $i'$, $z'$, overlayed on the AKARI NEP-Deep 
false-colour image (dark-black background).  \label{fig:maruma2}}
\end{figure}

Sky coverage of the multi-wavelength data around NEP is shown in Figure~\ref{fig:maruma1}.
A zoom-up view around the NEP-Deep survey area with areal coverage of the new optical-NIR images is given
in Figure~\ref{fig:maruma2}.
A list of currently available multiwavelength data is shown in Table~\ref{tab:maruma1}.
Note that the list focuses on the NEP-Deep although the survey area of some data covers
the NEP-Wide as well. 
The data with major updates are shown in boldface. The quality of the
nine AKARI/IRC band images has been greatly improved and a new band-merged catalogue was 
created with improved depth ($\sim20\%$) and reliability~\citep{murata2013}. The 
catalogue was opened to public in October 2013 through the ISAS DARTS archive. As for
the optical-NIR data, new band-merged catalogue ($u^{*}$, $g'$, $r'$, $i'$, $z'$, $Y$,
$J$, $K_{\rm s}$) was generated from newly obtained deep images with CFHT/MegaCam and
WIRCam~\citep{oi2014}. The 300~ks Chandra X-ray image data are also published in \citet{krumpe2014}.
 Regarding the FIR/submm data Herschel/PACS observations could
be undertaken just before the running out of the cryogen for Herschel. The imaging data
analysis for both PACS and SPIRE has been done to some level (see the paper by
Pearson et al. in this proceeding). Significant progress was also seen in the spectroscopic
follow-up; Keck/DEIMOS optical spectroscopic observations were undertaken for $\sim$1000 
sources, and also a GTC/OSIRIS-MOS observing run was successful. NIR (1.0-1.8~$\mu$m) 
data with Subaru/FMOS were also successful and good spectra were obtained for $\sim$100
sources. 

\begin{table*}[t]
\caption{The NEP-Deep Multi-wavelength data. Recent (last two years) progress are highlighted 
in boldface.\label{tab:maruma1}}
\centering
\begin{tabular}{lrrrc}
\toprule
Observatory/Instrument  &  Band/Filter              & Sensitivity           & Area/Target        &  Status (Sep. 2014)            \\
\midrule
AKARI/IRC               & 2.4-24~$\mu$m, 9 bands    & 90$\mu$Jy@15$\mu$m    &  0.5 deg$^2$       &  {\bf updated, published}$^1$  \\     
AKARI/IRC (Spec.)       & 2.4-12~$\mu$m, 9 bands    & $\sim$1mJy@9~$\mu$m   &  $\sim$100 sources &  paper in prep.                \\    
Subaru/SuprimeCam       &  $BVRi'z'$, NB711         & $B$ =28 ABmag         & $27'\times34'$     &  paper in prep.                \\
Subaru/{\bf HSC}        & $r$                       & $r$ =27.2 ABmag       & 5.4 deg$^2$        &  analysis on-going             \\
Subaru/FOCAS            &  optical spec.            & $R\sim 24$ ABmag      & 57 sources         &  paper in prep.                \\
Subaru/{\bf FMOS}       &  $JH$ spec.               & $J\sim 19$ ABmag      & $\sim$700 sources  &  paper in prep.                \\
Keck/DEIMOS             & opt. spec.(July 2008)     & $R\sim 24$ ABmag      & 420 sources        & analysis completed             \\
                        & opt. spec.(July 2011)     &                       & $\sim$600 soures   & analysis on-going              \\
                        & opt. spec.{\bf(Aug. 2014)}  &                     & $\sim$200 soures   & analysis on-going              \\
MMT,WIYN                & optical spec.             &                       & 5.4 deg$^2$        & {\bf published}$^2$          \\
{\bf GTC/OSIRIS-MOS}    & optical spec.             &                       & 190 sources        & analysis on-going             \\ 
CFHT/MegaCam            & $g'$ $r'$ $i'$ $z'$       &  $r' \sim 25$ ABmag   &  2~deg$^2$         &  published$^3$                   \\
                     & $u^{*}$ $g'$ $r'$ $i'$ $z'$  & $r' \sim 26.5$ ABmag  &  1~deg$^2$         &  {\bf published}$^4$            \\
CFHT/WIRCam             & $YJK_{\rm s}$             & $K_{\rm s}\sim 24$ ABmag &  0.5~deg$^2$    &  {\bf published}$^4$            \\
KPNO 2.1m/Flamingos     & $JK_{\rm s}$              & $K_{\rm s}$=20 Vega mag  & $25'\times30'$  & published$^5$                  \\
KPNO 4m/NEWFIRM         & $HK_{\rm s}$              & $K_{\rm s}$=22 ABmag  & $27'\times27'$     & analysis completed             \\
Chandra/ACIS-I          & 0.5-7~keV                 & (30-50ksec)           &  0.34 deg$^2$      & {\bf published}$^6$            \\    
GALEX                   & NUV, FUV                  & NUV$\sim$26 AB mag    & circular, 1.0 deg$\phi$  &    paper in prep.   \\
Herschel/SPIRE          & 250, 350, 500~$\mu$m       & $\sim$10~mJy          & 7.1~deg$^2$        & {\bf paper in prep.}        \\
$Herschel/PACS $        & 100, 160~$\mu$m           & 5-10~mJy              &  0.5~deg$^2$       & {\bf paper in prep.}        \\
{\bf JCMT/SCUBA-2}      & 450, 850~$\mu$m           & $\sim$ 1mJy           &  0.25~deg$^2$      & analysis on-going               \\
WSRT                    & 1.5GHz                    & 0.1~mJy               & $\sim$1.7 deg$^2$  &  published$^7$                   \\
GMRT                    & 610MHz                    &                       & $\sim$0.5 deg$^2$  & analysis on-going              \\
\bottomrule
\end{tabular}
\tabnote{$^1$~\citet{murata2013}; $^2$~\citet{shim2013}; $^3$~\citet{hwang2007}; 
$^4$~\citet{oi2014}; $^5$\citet{imai2007}; $^6$~\citet{krumpe2014};  $^7$~\citet{white2010} }
\end{table*}

\section{Recent Scientific Progress\label{sec:outcome}}

Great progress was seen in the study of dusty star-formation and AGN activity 
out to $z$=2 by mainly utilizing the unique AKARI/IRC photometry data
covering continously 2.4-24~$\mu$m wavelengths. \citet{hanami2012} showed that
the rest-frame 8~$\mu$m and 5~$\mu$m luminosities are good tracers of star-forming
and AGN activities from their Polycyclic Aromatic Hydrocarbons (PAH) and dusty tori
emissions, respectively. As for
the AGN dominated MIR-selected sources (inferred from their MIR SEDs), \citet{krumpe2014}
found a high ($\sim 40\%$) X-ray detection rate, while sources without any sign of AGN 
activity in their MIR SEDs have a very low X-ray detection rate of 3\%. 
They also concluded
that roughly 30\% of IR-selected AGN are strong Compton-thick AGN candidates; this is about
to be verified by rest-frame X-ray stacking (Miyaji et al. in prep.). On the other
hand, \citet{murata2014} extracted the pure starburst sources after excluding the AGN 
candidates by the SED fitting. They found that rest-frame 8~$\mu$m / 5~$\mu$m luminosity ratio
(e.g. a proxy of PAH equivallent width) is not proportional to the starburstiness 
(specific star-formation rate (sSFR) normalized by that of the main-sequence)
at higher starburstiness, indicating the PAH feature deficit under the intense starburst.

Many on-going projects will lead to publications in the near future. NIR spectroscopic follow-up
of MIR selected sources provides the opportunity to investigate metallicity of $z \sim 0.8$
dusty IR luminous galaxies by [\ion{N}{ii}]/$H_{\alpha}$ ratio~(Oi et al., in prep.). 
The sSFR and dust attenuation evolution can be studied by using the rest-frame 8~$\mu$m selected 
sample out to $z$=2 (see Buat et al. paper in this conference), where sSFR and dust attenuation
are derived by the 
SED fitting with CIGALE (Code Investigating  GALaxy Emission, http://cigale.lam.fr).
It is notable that (U)LIRGs' SED characteristics is under careful investigation using CIGALE 
fitting (Malek et al. in prep.).
By using the New FIR (PACS) photometry data re-analysis of sSFR of the MIR-selected
galaxies, and of AGN fraction by also using the X-ray (Chandra) data is on-going (Ishigaki et al.
in prep.).
By using 24~$\mu$m selected galaxies at different redshift, evolution of clustering is
examined (Solaz et al. 2014, submitted), and stellar mass dependence of sSFR  between 
NEP (AKARI) and SXDF (Spitzer) is examined  (Fujishiro et al. paper in this conference).

\section{Summary and Future Prospects}

The recent updates of the NEP-Deep multi-wavelength survey covering from X-ray 
to radio-wave is presented. For existing data, a few catalogue papers were published,
while new multiwavelength surveys (Subaru/HSC, JCMT/SCUBA-2) have been undertaken.
Significant progress was seen in the dusty AGN/starburst classification (or 
determination of AGN fraction for each MIR-selected source), and determination of 
radiation hardness or star-formation mode by using the UV-submm SED. These outcomes
are useful to perform the new goal of the project, `to resolve the nature of the CSF 
 history at the violent epoch, and to find a clue to understand
its decline from $z$=1 to present universe.' In the near future, we aim to obtain
multiband photometry data with Subaru/HSC (and CFHT/MegaCam $u^{*}$) over 5.4~deg$^2$ 
NEP-Wide survey area in order to significantly increase the number of (U)LIRGs with 
accurate redshift, sSFR, and AGN fraction. 

It is also noteworthy that the NEP is the legacy field thanks to its high visibility
by the space observatories, such as eROSITA, Euclid, JWST, and SPICA. SPICA, the next
generation 3~m class cooled space telescope is extremely powerful to study the rise
and fall of the CSF in the universe via the PAH equivallent width diagnostics (see
Wada et al. paper in this conference).

%\section{The First Section}
%\subsection{The First Subsection}
%\subsubsection{The First Subsubsection}
%
%Some text, showing the ion \ion{H}{ii}, an example footnote,\footnote{Example footnote} and an example reference to \citet{einstein1905}.

%%% ACKNOWLEDGMENTS (IF ANY) %%%%%%%%%%%%%%%%%%%%%%%%%%%%%%%%%%%%%%%%

\acknowledgments

We would like to thank all $AKARI$ team members for their support on this 
project. 
The AKARI NEP-Deep survey project activities are mainly supported by a JSPS grant
23244040, and also partly supported by the Chandra Guest Observer
 support GO1-12178X, CONACyT grant 83564, and the Institute of Space and 
Astronautical Science, Japan Aerospace Exploration Agency.

%%% APPENDICES (IF ANY) %%%%%%%%%%%%%%%%%%%%%%%%%%%%%%%%%%%%%%%%%%%%%
%
%\appendix
%\section{Appendix Title}
%
%Some text.
%
%%% CALL LIST OF REFERENCES (natbib STYLE) %%%%%%%%%%%%%%%%%%%%%%%%%%

\end{document}